\documentclass[secnumarabic,amssymb, nobibnotes, aps, prl]{revtex4-2}

\usepackage{harpoon}
\usepackage{graphicx}
\usepackage{lipsum, babel}
\begin{document}

\title{Solid-like high harmonic generation from rotationally periodic systems}%

\author{Yigeng Peng}
\affiliation{Institute of Ultrafast Optical Physics, Department of Applied Physics \& MIIT Key Laboratory of Semiconductor Microstructure and Quantum Sensing, Nanjing University of Science and Technology, Nanjing 210094, P R China}

\author{Tong Wu}
\affiliation{Institute of Ultrafast Optical Physics, Department of Applied Physics \& MIIT Key Laboratory of Semiconductor Microstructure and Quantum Sensing, Nanjing University of Science and Technology, Nanjing 210094, P R China}

\author{Guanglu Yuan}
\affiliation{Institute of Ultrafast Optical Physics, Department of Applied Physics \& MIIT Key Laboratory of Semiconductor Microstructure and Quantum Sensing, Nanjing University of Science and Technology, Nanjing 210094, P R China}

\author{Lihan Chi}
\affiliation{Institute of Ultrafast Optical Physics, Department of Applied Physics \& MIIT Key Laboratory of Semiconductor Microstructure and Quantum Sensing, Nanjing University of Science and Technology, Nanjing 210094, P R China}

\author{Chao Yu}
\email[Corresponding author :]{chaoyu@njust.edu.cn}
\affiliation{Institute of Ultrafast Optical Physics, Department of Applied Physics \& MIIT Key Laboratory of Semiconductor Microstructure and Quantum Sensing, Nanjing University of Science and Technology, Nanjing 210094, P R China}

\author{Ruifeng Lu}%
\email[Corresponding author :]{rflu@njust.edu.cn}
\affiliation{Institute of Ultrafast Optical Physics, Department of Applied Physics \& MIIT Key Laboratory of Semiconductor Microstructure and Quantum Sensing, Nanjing University of Science and Technology, Nanjing 210094, P R China}
\date{\today}%

\begin{abstract}
	High harmonic generation (HHG) from crystals in strong laser fields has been understood by the band theory of solid, which is based on the periodic boundary condition (PBC) of translational invariant. For systems having PBC of rotational invariant, in principles an analogous Bloch theorem can be developed and applied. Taking a ring-type cluster of cyclo[18]carbon as a representative, we theoretically suggest a $quasi$-band model and study its HHG by solving time-dependent Liouville–von Neumann equation. Under the irradiation of circularly polarized laser, explicit selection rules for left-handed and right-handed harmonics are observed, while in linearly polarized laser field, cyclo[18]carbon exhibits solid-like HHG originated from intra-band oscillations and inter-band transitions, which in turn is promising to optically detect the symmetry and geometry of controversial structures. In a sense, this work presents a connection linking the high harmonics of gases and solids. 
\end{abstract}

\maketitle

In past decades, high harmonic generation (HHG) of gaseous media with the interaction of a driving strong laser has been extensively studied for its potential applications in optical technology, such as attosecond pulse generation and spectral characterization of structural or electronic dynamics \cite{Paul2001}. At the meantime, the “ionization-acceleration-recombination” three-step model \cite{Corkum1993} has been widely recognized. Limited by the density of gaseous targets, the harmonic conversion efficiency is dissatisfactory for widespread usage. Because of the high density of atoms in solids, efficient high harmonics from crystalline bulk and two-dimensional materials have been observed in recent years \cite{Luu2015,Liu2017}. In this emerging field, many interesting phenomena have been reported, including intra-band Bloch oscillation \cite{Luu2015}, inter-band interference \cite{Garg2016}, atomic-like HHG \cite{Tancogne-Dejean2018,Yu2020}, and even-order harmonics due to symmetry breaking of crystal structure \cite{Jiang2018,Jiang2019} etc.

A cluster which is normally composed of several up to tens of thousands of atoms, could be regarded as the intermediate between atom and solid material that presents both atomic and solid characters. In fact, the HHG properties of noble gas clusters are found quite different with the growth of cluster size \cite{Ruf2013,Park2014}. However, all previous HHG studies of clusters are based on the three-step model, basically similar to atomic systems. An electron is first ionized and accelerated by the laser field, then returns and recombines with the mother cation when the electric field changes sign, finally emits a high-energy photon. In comparison with atoms, the enhancement of HHG from clusters is due to that the delocalized electronic wave function leads to larger polarizability and higher recollision possibility. Please keep in mind that the HHG mechanism starts with ionization in these works. 

For solid crystals, electron delocalization in lattice structure can be described by the periodic boundary condition (PBC) of the translational invariant. As shown in Fig. 1, another kind of the PBC can be satisfied from the rotational invariant. For ring-type systems of $N$-fold rotational symmetry, similar to the translation invariance in ordered solids, their eigenfunctions satisfy a generalized Bloch theorem \cite{Lee2009,Kit2011,Dobardzic2015} which is suggested as:
\begin{eqnarray}
	\label{RotBloch}
	\psi_J(\theta+\theta_0)=\psi_J(\theta)\cdot e^{iJ\theta}
\end{eqnarray}
where $\theta_0=2\pi/N$ represents the rotation angle that makes the system rotational invariant. $J$ should be integer according to the boundary condition: $\psi_J (\theta+2\pi)=\psi_J(\theta)$ and $\psi_J =\psi_{J+N}$, which means that only $N$ independent values of $J$ are available. Eq. (1) enables the electron delocalization in ring-type systems. Analogous to crystalline solids, the physical quantity $J$ here can be regard as some kind of angular momentum due to the rotational invariability by certain angles ($quasi$-angular momentum, QAM), which is similar to the crystal momentum of solids due to the translational invariability by certain vectors. Meanwhile, $N$ can be regarded as some kind of reciprocal lattice vector, and the region of $J$ values composes an analogical Brillouin zone ($quasi$-reciprocal space). Then a $quasi$-energy-band structure can be obtained by combining the eigen-energies of all $J$, as shown in Fig. 2(a). Considering the physical analogy between crystals with translational PBC and ring-type systems with rotational PBC, in this work we propose a solid-like HHG mechanism of ring-type cluster cyclo[18]carbon in the framework of generalized Bloch theorem with rotational PBC.

\begin{figure}[htbp]
	\centering
	\label{Blo}
	\includegraphics[width=0.5\textwidth]{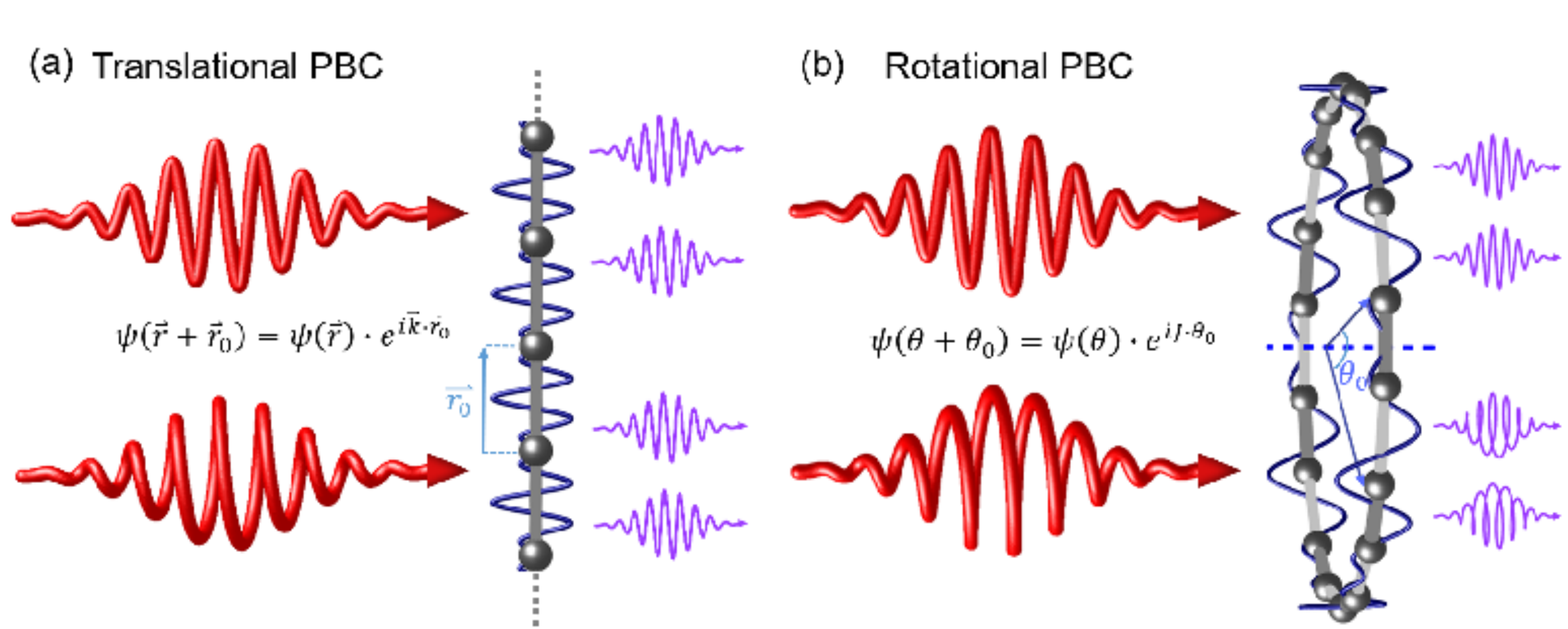}
	\caption{Schematic of the generalized Bloch theorem for the systems with (a) translational PBC and (b) rotational PBC.}
\end{figure}

The cyclo[18]carbon is an all-carbon atomic ring composed of 18 carbon atoms. Since the earliest study in the 1960s by Hoffmann \cite{Hoffmann1966}, its physical properties have been intensively explored including geometrical structure \cite{Parasuk1991} and electron correlation \cite{Torelli2000}. Very recently, breakthroughs in synthesis and characterization of cyclo[18]carbon have been achieved by Kaiser et al. \cite{Kaiser2019,Scriven2020}, afterward great interests have been aroused because of the unique properties and potential applications of this cluster. In fact, there are two possible geometric structures of cyclo[18]carbon: (i) a cumulenic structure with C-C bonds of equal length has a $D_{18\rm{h}}$ symmetry, and (ii) a polyynic structure with alternating triple and single bonds shows a $D_{9\rm{h}}$ symmetry. The experiment \cite{Kaiser2019} using scanning tunneling microscopy (STM) and atomic force microscopy (AFM) suggested the latter, which may display interesting non-linear optical effects due to its $\pi$ electron delocalization. Based on the experimental findings, we focused on the polyynic structure (i.e. $N = 9$), and the integer values of $J$ are from -4 to 4. Quantum chemistry calculations for electronic structures were carried out by Gaussian 16 package \cite{g16}, and M06-2X functional with basis set of 6-311G++(d,p) was employed. The geometry of polyynic cyclo[18]carbon was optimized with the alternating bond lengths of 1.226$\mathring{A}$ and 1.346$\mathring{A}$ and all 18 C-C-C angles fixed at 160º, in good agreement with the reported values \cite{Baryshnikov2019}. The molecular orbits (MOs) were analyzed with the Multiwfn package \cite{Lu2012}. The calculated gap between the highest occupied MO and the lowest unoccupied MO by a correction method \cite{Li2020} is 2.72 eV, implying that the polyynic cyclo[18]carbon is a semiconductor.

As is well understood, polyynic cyclo[18]carbon consists of two sets of 18-center delocalized $\pi$ electrons (called $\pi$-in and $\pi$-out) with 18 occupied MOs and 18 unoccupied ones. In order to satisfy Eq. (1) perfectly, we perform a linear transformation for the energy degenerated MOs as $\psi_{\pm J}=\varphi_J^a\pm i \varphi_J^b$, where $\varphi_J^a $and $\varphi_J^b$ represent a couple of degenerated $\pi$-in (or $\pi$-out) MOs ( the isosurfaces of these MOs are shown in suppl Fig. s1). We divide 36 $\pi$-in and $\pi$-out MOs into 4 bands according to their energies and corresponding $J$ values. The $quasi$-energy band structures are presented in Fig. 2(a) and Fig. s2. At the same $J$ values, the occupied $\pi$-in and $\pi$-out bands are almost degenerated in energy while significant energy differences exist between the unoccupied ones. Here we can regard the occupied band as the $quasi$-valence band ($quasi$-VB) and the unoccupied one as the $quasi$-conduction band ($quasi$-CB), thus the four bands are named as $quasi$-VB$_1$, $quasi$-CB$_1$, $quasi$-VB$_2$, $quasi$-CB$_2$, in which the $\pi$-in bands are subscribed by 1 and $\pi$-out bands are subscribed by 2. Owning to the nearly equivalence between $\pi$-in and $\pi$-out VBs, in Fig. 2, we only present the properties of the $\pi$-in bands for clarity.
\begin{figure}[htbp]
	\centering
	\label{Dispersion}
	\includegraphics[width=0.5\textwidth]{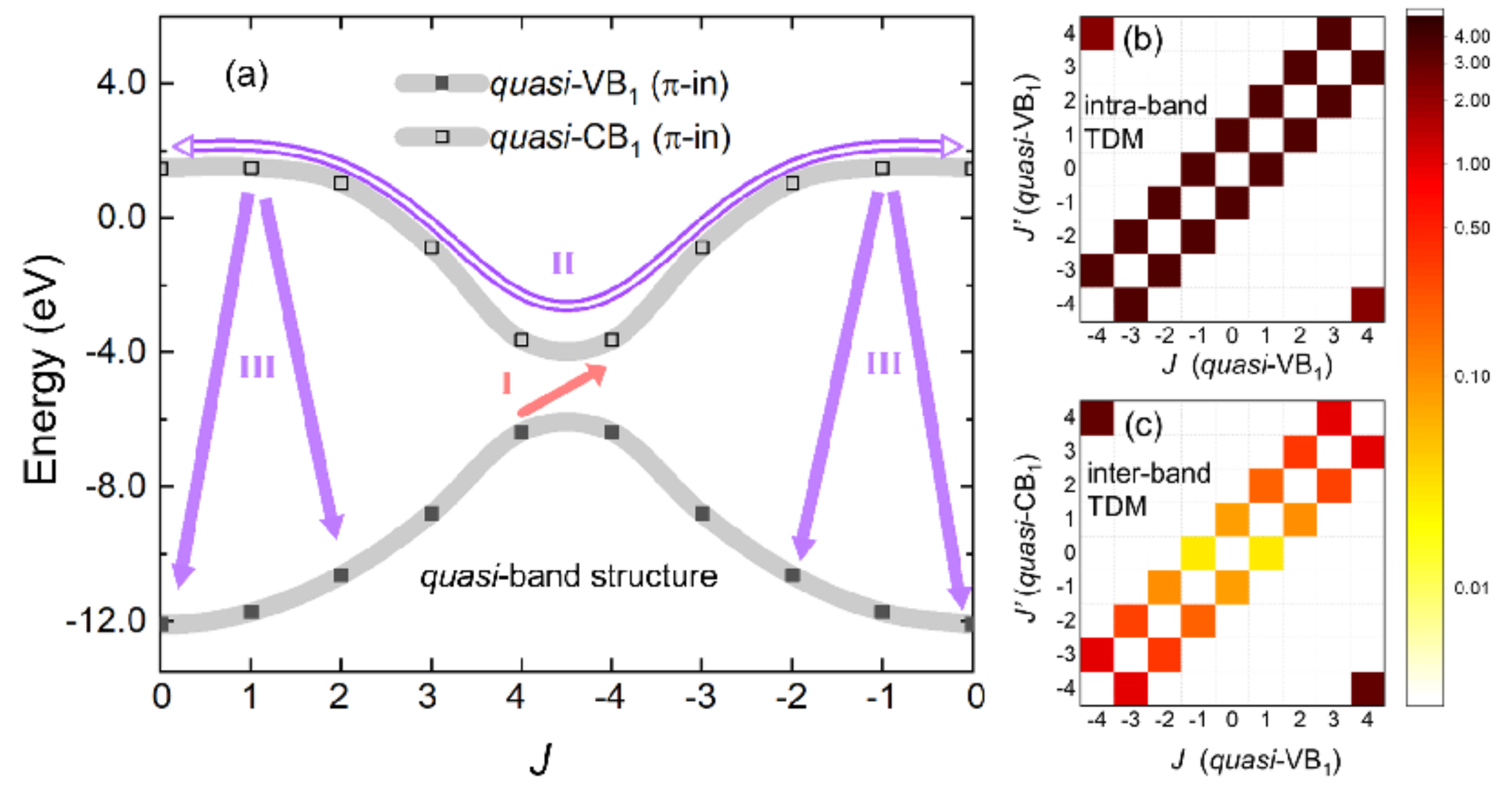}
	\caption{(a) $Quasi$-band structure of polyynic cyclo[18]carbon in the framework of generalized Bloch theorem with rotational PBC, including inter-band transitions (I, III; solid arrows) and intra-band oscillations (II; hollow arrows) responsible for solid-like HHG. The absolute values of (b) intra-band TDM for $quasi$-CB$_1$ (the values for $quasi$-VB$_1$ are almost the same as $quasi$-CB$_1$) and (c) inter-band TDM between $quasi$-VB$_1$ and $quasi$-CB$_1$. }
\end{figure}

According to Eq.\ref{RotBloch}, the transition dipole moment (TDM) between occupied band and unoccupied band pair for different $J$ reads as 
\begin{eqnarray}
	\overrightharp{D}_{nn'}^{JJ'}=\left\langle\psi_{nJ}\right|\overrightharp{r}\left|\psi_{n'J'}\right\rangle
\end{eqnarray}
which satisfies the selection rule $J-J'=\pm 1$ for ensuring the conservation of angular momentum, and $n/n'$ represents the band index. The calculated values for $\pi$-in bands are shown in Figs. 2(b) and 2(c). The intra-band TDM values nearby the diagonal squares in Fig. 2(b) are close to 3.5 a.u., corresponding to half of the radius of cyclo[18]carbon as $\left\langle\psi_{nJ}\right|x\left|\psi_{n'J'}\right\rangle\approx1/2\pi\int_{0}^{2\pi}e^{-iJ\theta}R_{\rm{c}18}\cos(\theta) e^{iJ\theta}d\theta=R_{\rm{c}18}/2$, where $R_{\rm{c}18}\approx 7~a.u.$ This property is also valid for larger rings such as cylco[22]carbon, cyclo[26]carbon, etc. In Figs. 2(b) and 2(c), solid squares at the corners should be noticed, implying that the transition between $J = -4$ and $J = 4$ is allowed because of the rotational periodic property. 

To investigate the interaction between cyclo[18]carbon and external laser field, we numerically solve the time-dependent Liouville–von Neumann equation:
\begin{eqnarray}
	i\frac{\partial \rho}{\partial t}=\left[H,\rho\right]-\frac{\rho^d-\rho_0^d}{T_1}-\frac{\rho-\rho^d}{T_2}
\end{eqnarray}
where $\rho$ represents the density matrix, $\rho^d$ and $\rho_0^d$ are the diagonal term of $\rho$ and the initial density matrix $\rho_0$, respectively. $T_1$ is the relaxation time and $T_2$ is the dephasing time. The Hamiltonian matrix is $H=U_n^J+ \overrightharp{E}(t)\cdot \overrightharp{D}_{nn'}^{JJ'}$, where diagonal $U_n^J$ is the $quasi$-energy band as shown in Fig. 2(a) and Fig. s2, $\overrightharp{D}_{nn'}^{JJ'}$ is the abovementioned TDM, and $\overrightharp{E}(t)$ represents time-dependent electric field of the driving laser.

The time-dependent induced dipole $\overrightharp{P}(t)=Tr(\rho(t) \overrightharp{D} )$ can be obtained through the density matrix, then the HHG can be calculated from the Fourier transform of the dipole acceleration $\overrightharp{a}(t)=\ddot{\overrightharp{P}}(t)$. In this work, we consider both circularly- and linearly-polarized lasers in the ring plane, and use the laser parameters with wavelength of 1600 nm, peak intensity of 1.0$\times10^{13} W/cm^2$, and full-width at the half maximum of 45 $fs$. 

There have been several strict theoretical proofs \cite{Liu2016,Fleischer2014,Ceccherini2001,Alon1998} of the HHG selection rule for the N-fold system induced by circularly-polarized laser, and the $kN\pm1 $(the integer $k\geq1$) harmonics are found. Such a selection rule is also valid in the polyynic cyclo[18]carbon of 9-fold symmetry. As shown in Fig. 3(a), the 10th, 19th and 28th harmonics are right-handed following the same polarization of the incident light while the 8th, 17th and 26th ones are left-handed. Here, using the QAM image, the selection rule can be visually interpreted. As shown in Fig. 3(b), a circularly-polarized photon (red circled arrows) carries one $\hbar$ angular momentum and the $n$th harmonic generation takes place when the system absorbs $n$ photons (i.e., gets $n\hbar$ angular momentum), and transits back to the initial state by emitting a photon with $\pm\hbar$ angular momentum (orange and green bended arrows). Such processes lead to $(n\pm1)\hbar$ angular momentum changes on the systems, so the HHG induced by circularly-polarized laser is not allowed for the systems of angular momentum conservation like atoms or linear molecules. For the systems with rotational symmetry but without the angular momentum conservation, the quantum number $J$ of their eigenfunctions in Eq. (1) which can be regarded as the QAM, is quite similar to the classical angular momentum, e.g., the $J$ value will increase by 1 after absorbing one circularly-polarized photon. For the studied ring-type systems, the $J$ value changes periodically according to Eq. (1). This makes the $J$ value unchanged after absorbing $N$ circularly-polarized photons, resulting in the $kN\pm1$ selection rule. 
\begin{figure}[htbp]
	\centering
	\label{Circular}
	\includegraphics[width=0.5\textwidth]{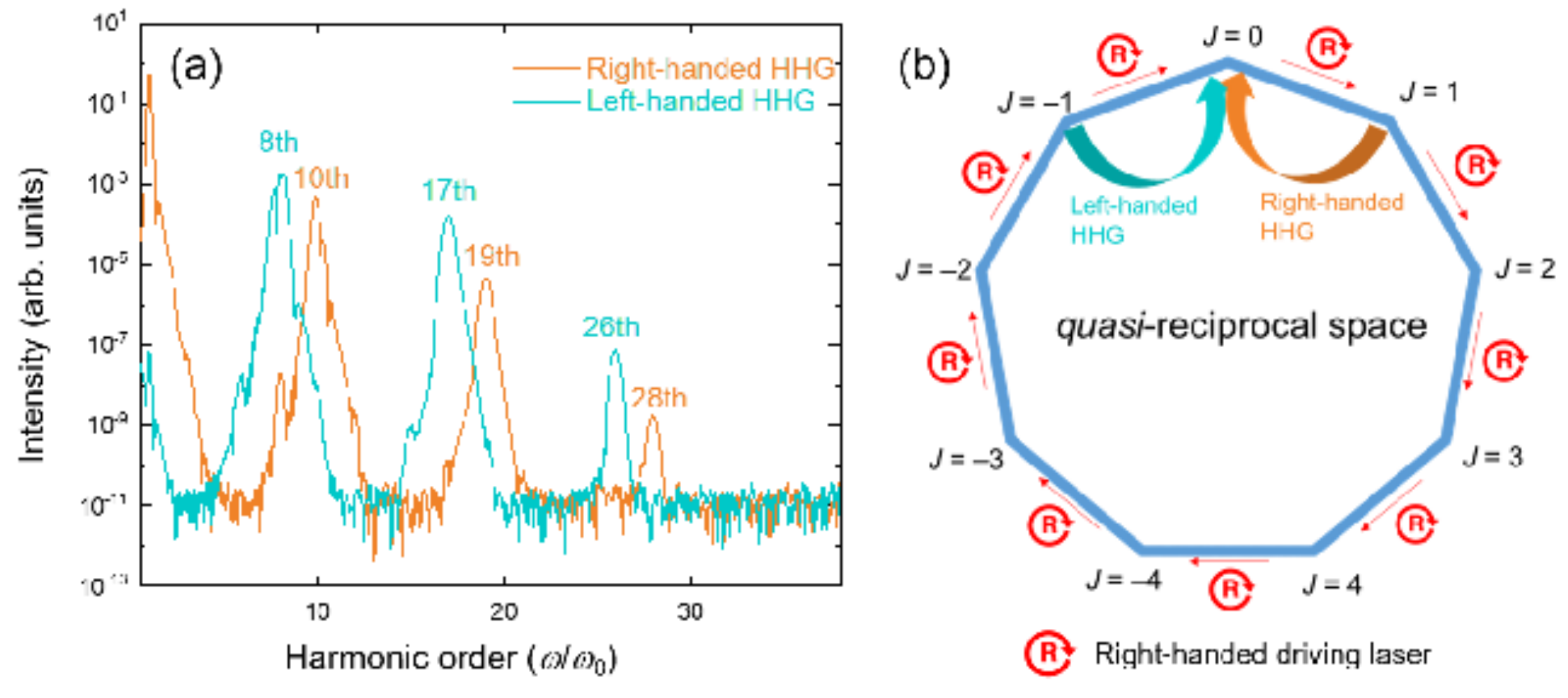}
	\caption{Simulated harmonic spectra of polyynic cyclo[18]carbon driven by right-handed circularly-polarized laser based on four bands ($quasi$-VB$_1$, $quasi$-CB$_1$, $quasi$-VB$_2$, $quasi$-CB$_2$) (left panel) and schematic of the selection rule in $quasi$-reciprocal space (right panel).}
\end{figure}

As is well known, the geometric structure of the cyclo[18]carbon has been subjected to long-term debates. Distinct from STM or AFM characterization of cyclo[18]carbon, here we put forward a nonlinear optical method to determine its structure by measuring the HHG induced by a right-handed light with the incident direction perpendicular to the ring plane, and the left-handed 8th and the right-handed 10th harmonics will appear at polyynic structure of $D_{9\rm{h}}$ symmetry, but disappear for cumulenic structure of $D_{18\rm{h}}$ symmetry (see suppl Fig. s6(a) for the left-handed 17th and the right-handed 19th harmonics of cumulenic structure).

Since polyynic cyclo[18]carbon ($D_{9\rm{h}}$) has its non-centrosymmetric axis, a linearly-polarized driving laser would lead to generation of both odd and even harmonics, which is dependent on the polarization direction of the incident light. As shown in Fig. 4(a), when the laser polarization is perpendicular to the symmetry axis, the polarization of the odd harmonics is parallel to incident direction while the polarization of the even ones is perpendicular to that of driving laser. When the incident polarization is parallel to the symmetry axis of the target (Fig. 4b), the polarization directions of both odd and even harmonics are parallel to laser polarization. On the contrary, cumulenic cyclo[18]carbon ($D_{18\rm{h}}$) is central inversion symmetric, and only odd order harmonics can be found with linearly polarized laser (see Figs. s6(b) and s6(c)). Such a special odd-even order harmonics generation phenomenon is consistent with that observed in other non-centrosymmetric systems with mirror axis like carbon monoxide \cite{Hu2017} and monolayer $MoS_2$ \cite{Liu2017,Liu2020}. Together with the aforementioned high-harmonic properties driven by circularly-polarized laser, we convincingly claim that high-order harmonics provide an efficient all-optical approach to detect the structural nuance between $sp$-hybridized and $sp^2$-hybridized carbon bonds. 
\begin{figure}[htbp]
	\centering
	\label{LinearHHG}
	\includegraphics[width=0.5\textwidth]{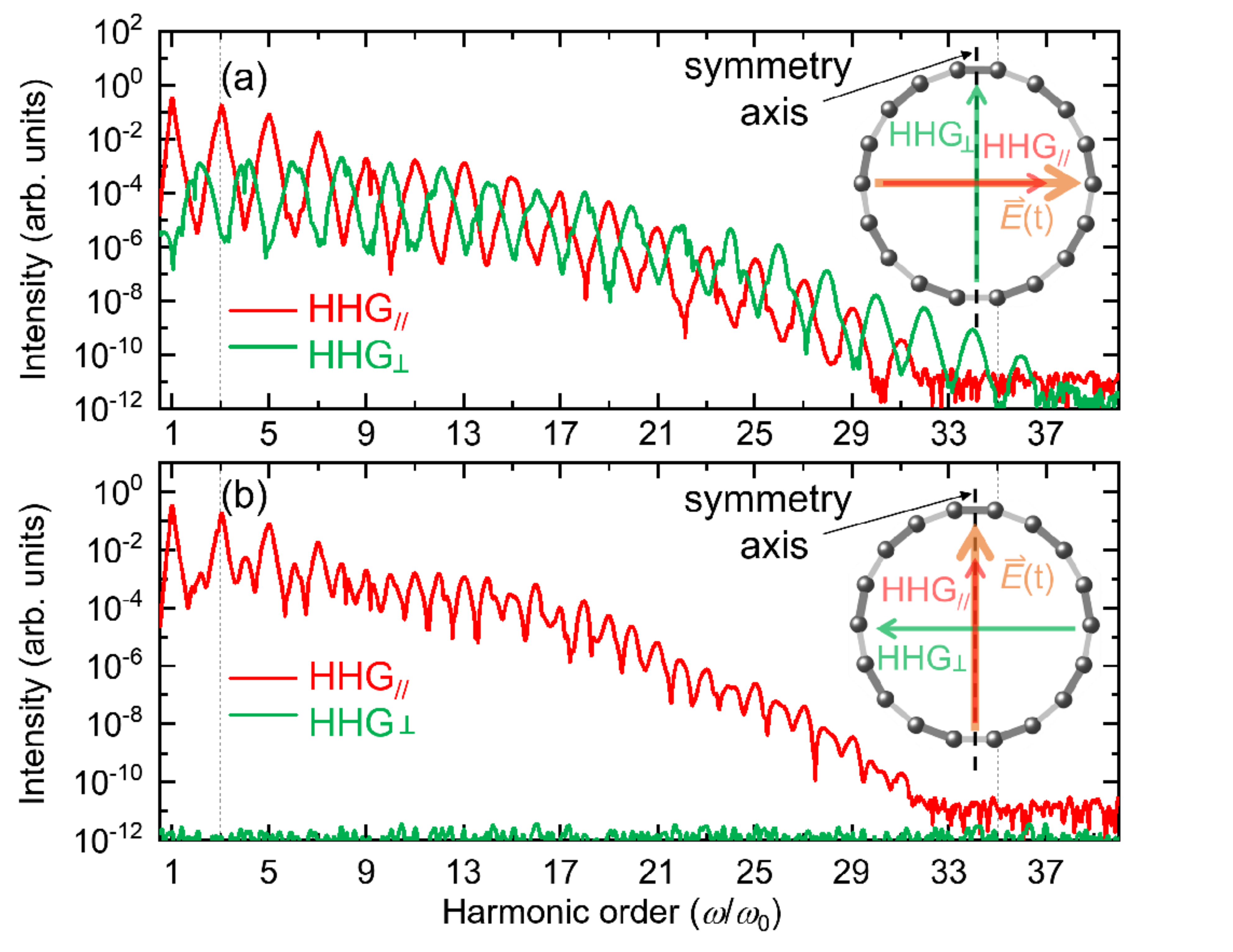}
	\caption{Simulated harmonic spectra of polyynic cyclo[18]carbon driven by linearly-polarized laser based on four bands. (a) is for the electric field of driving laser parallel to the symmetry axis and (b) is for the electric field of driving laser perpendicular to the symmetry axis. The HHG parallel to the incident laser (HHG$_{/\!/}$) and perpendicular to incident laser (HHG$_\perp$) are depicted. }
\end{figure}

From Figs. 2(b) and 2(c) and Tables s1-s6, in the beginning, electrons are excited via the transition from $quasi$-VB maximum to $quasi$-CB minimum, subsequently the main movement of the electrons under external laser field should be the intra-band oscillation due to the much larger intra-band TDM values, and the lower odd-order harmonics dominantly come from the intra-band oscillation. When the electrons transit back to the $quasi$-VB following the selection rule, the inter-band emission may happen and generate a photon with higher energy. Although the inter-band TDMs nearby the diagonal squares are much smaller, the higher orders in the harmonic spectrum are mainly originated from the inter-band processes. To verify this, we perform additional calculations by removing such small inter-band TDMs nearby the diagonal squares in Tables s5 and s6. The results shown in Figs. s3(a)-s3(c) show that the harmonic intensity at high orders (> 10) are obviously reduced.

In summary, we introduce a theory model to study the interaction between ultrafast strong laser and the systems satisfying rotational PBC. Taking cyclo[18]carbon as an example, we confirm its inter-band and intra-band mechanisms of HHG based on quasi energy bands. Since it is very similar to the HHG of solid crystals dependent on their band dispersions which can be theoretically simulated by solving semiconductor Bloch equations, we name it as solid-like HHG in the framework of generalized Bloch theorem. So far, although the theory of solid HHG have made rapid progress, it has not yet been fully developed. For the gaseous atoms or molecules, HHG is usually analyzed based on the three-step “ionization-acceleration-recollision” process. A similar “excitation-acceleration (Bloch oscillation)-recombination” process for solid HHG is based on band theory. Some pioneering studies \cite{Vampa2015nat,Vampa2015prl} have addressed this analogy clearly. Recent atomic-like HHG studies of two-dimensional materials \cite{Tancogne-Dejean2018} including one of our work \cite{Yu2020} provide a more direct comparison with the atomic systems. The solid-like HHG of ring-type clusters is from reverse thinking that has enriched the understanding of HHG field. Last but not the least, benefited from the clear physical pictures, we successfully justify the selection rule and odd-even features of HHG from 9-fold and 18-fold rotationally symmetric cyclo[18]carbon clusters, and a novel spectral characterization strategy using high-order harmonics is proposed to determine the suspicious structures. The theory and findings in this work are hopefully extended to other ring-type or ball-type molecules/clusters, nanotubes or artificial metamaterials with rotational PBCs that deserve much endeavor in the future.

\section{Acknowledgments}
This work was supported by NSF of China Grant (No. 11974185, 11904028, 12174195, and 11834004), the Natural Science Foundation of Jiangsu Province (Grant No. BK20170032).


%

\includegraphics[width=1.1\textwidth]{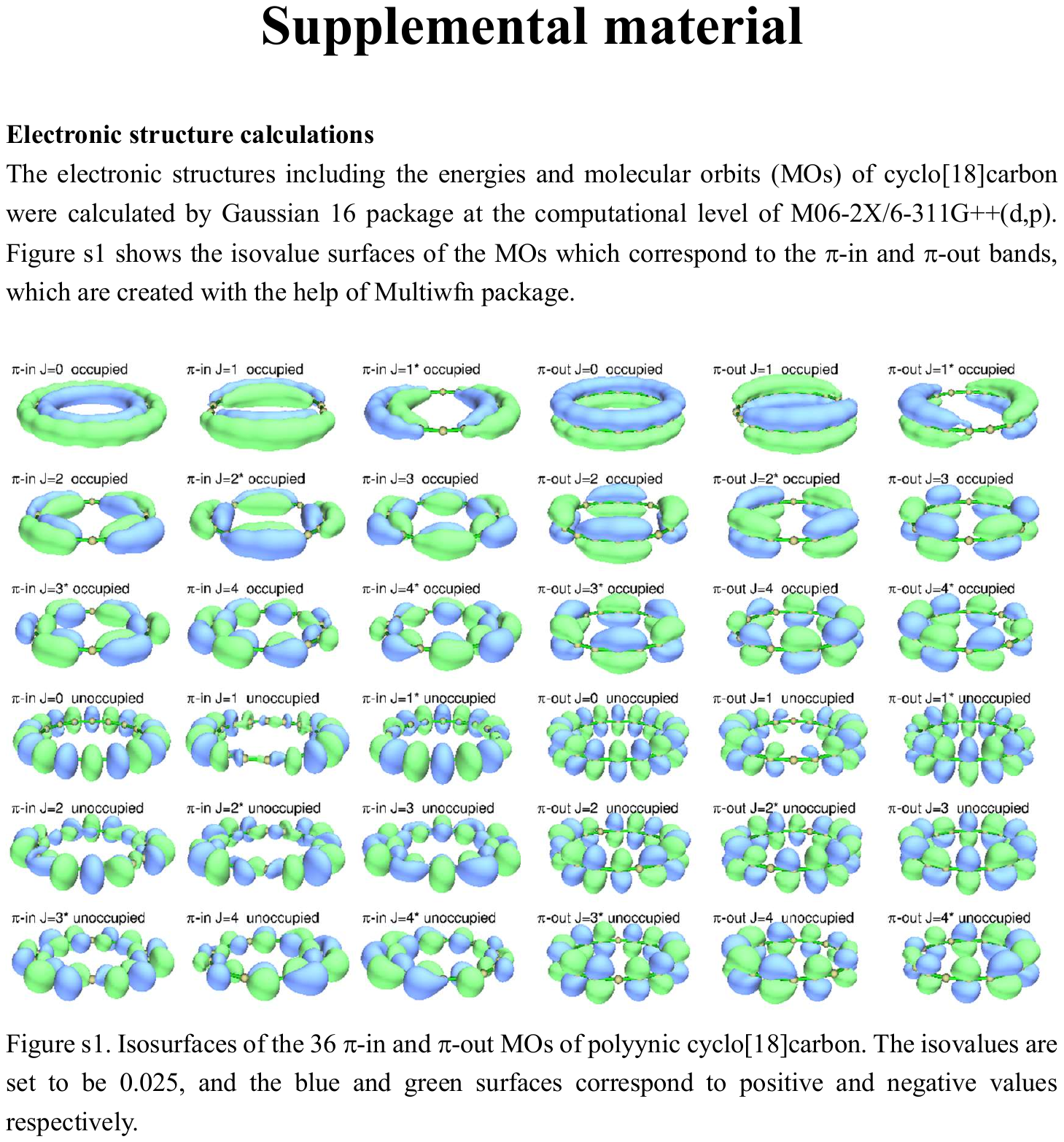}
\includegraphics[width=1.1\textwidth]{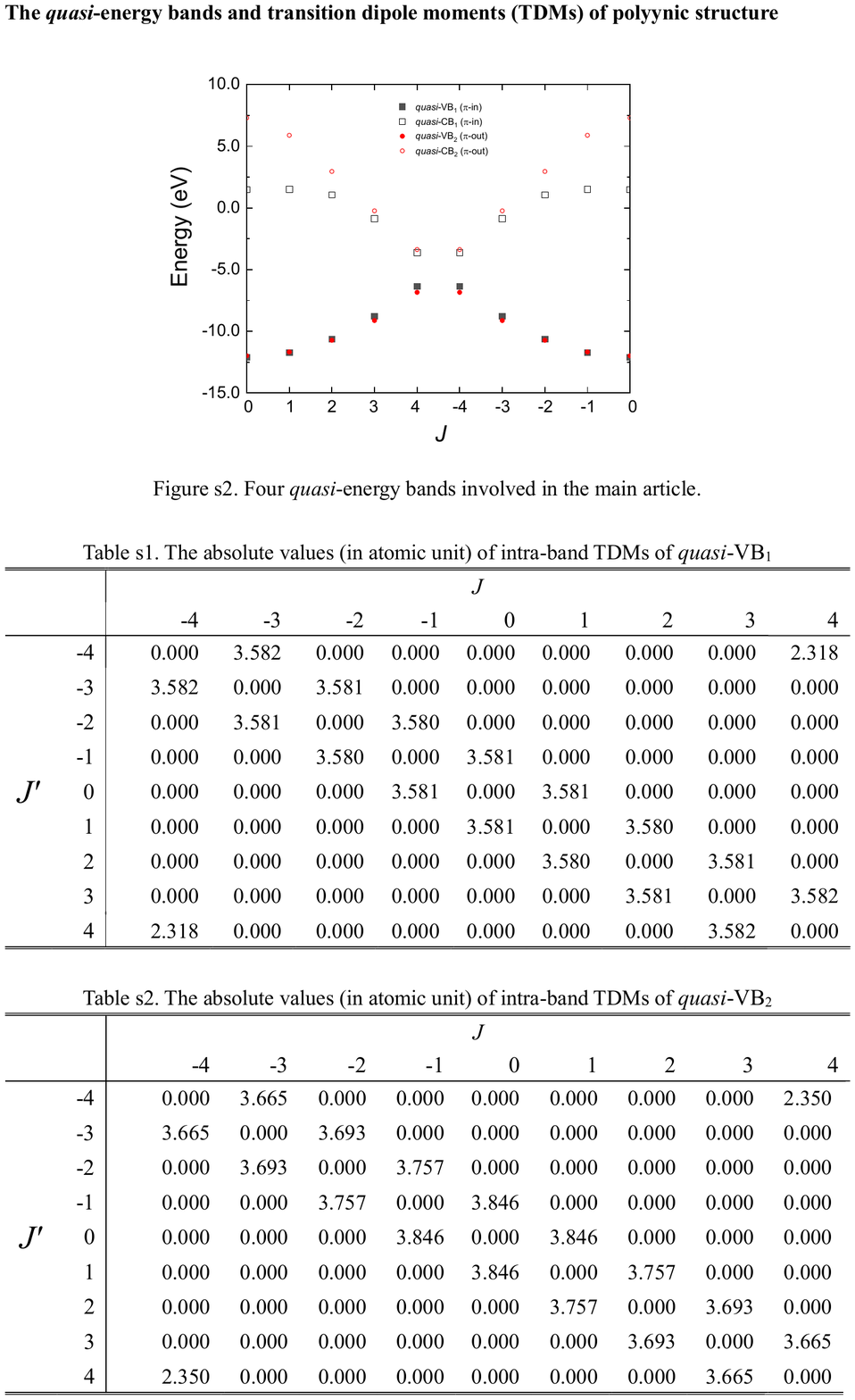}
\includegraphics[width=1.1\textwidth]{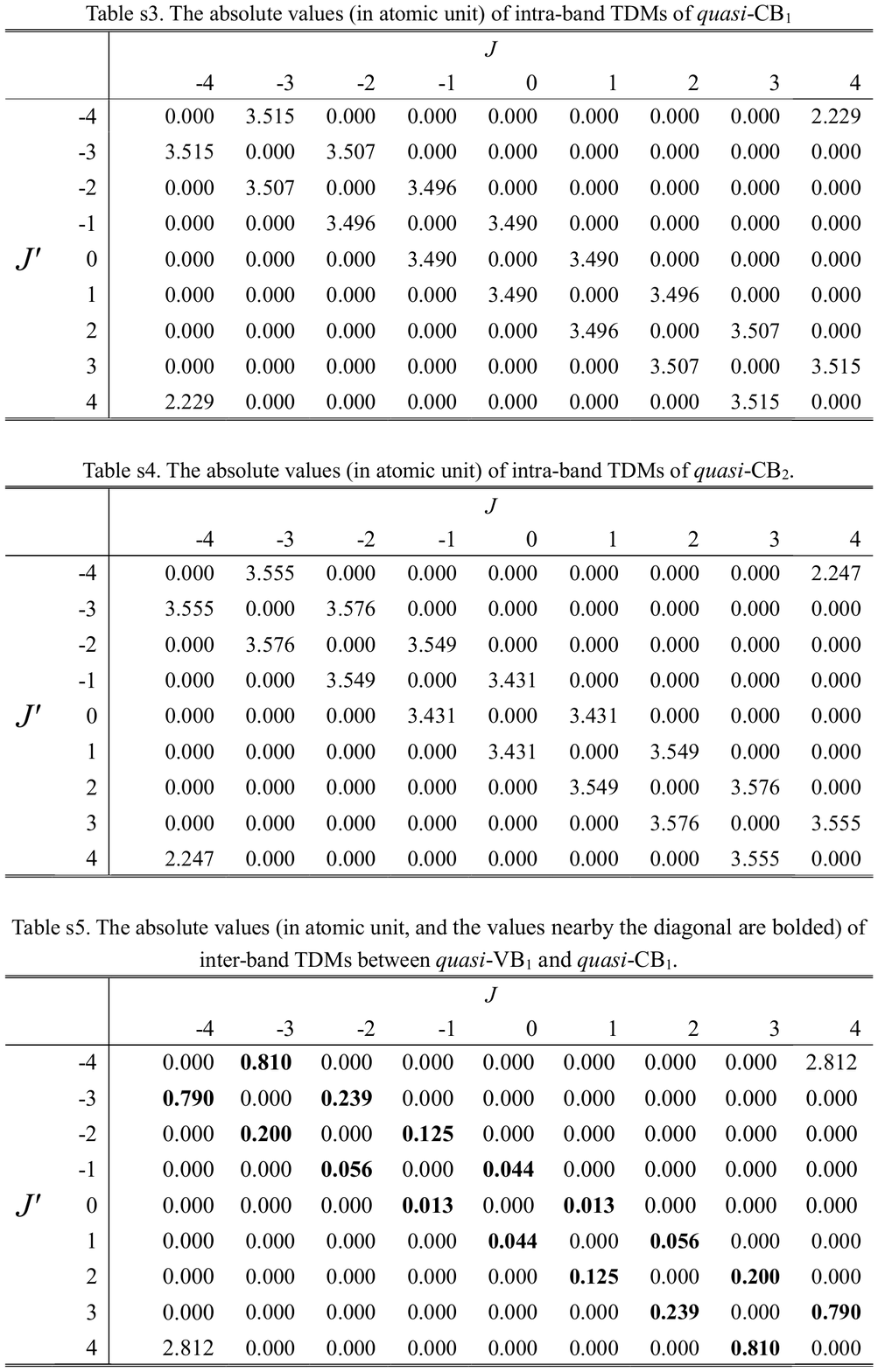}
\includegraphics[width=1.1\textwidth]{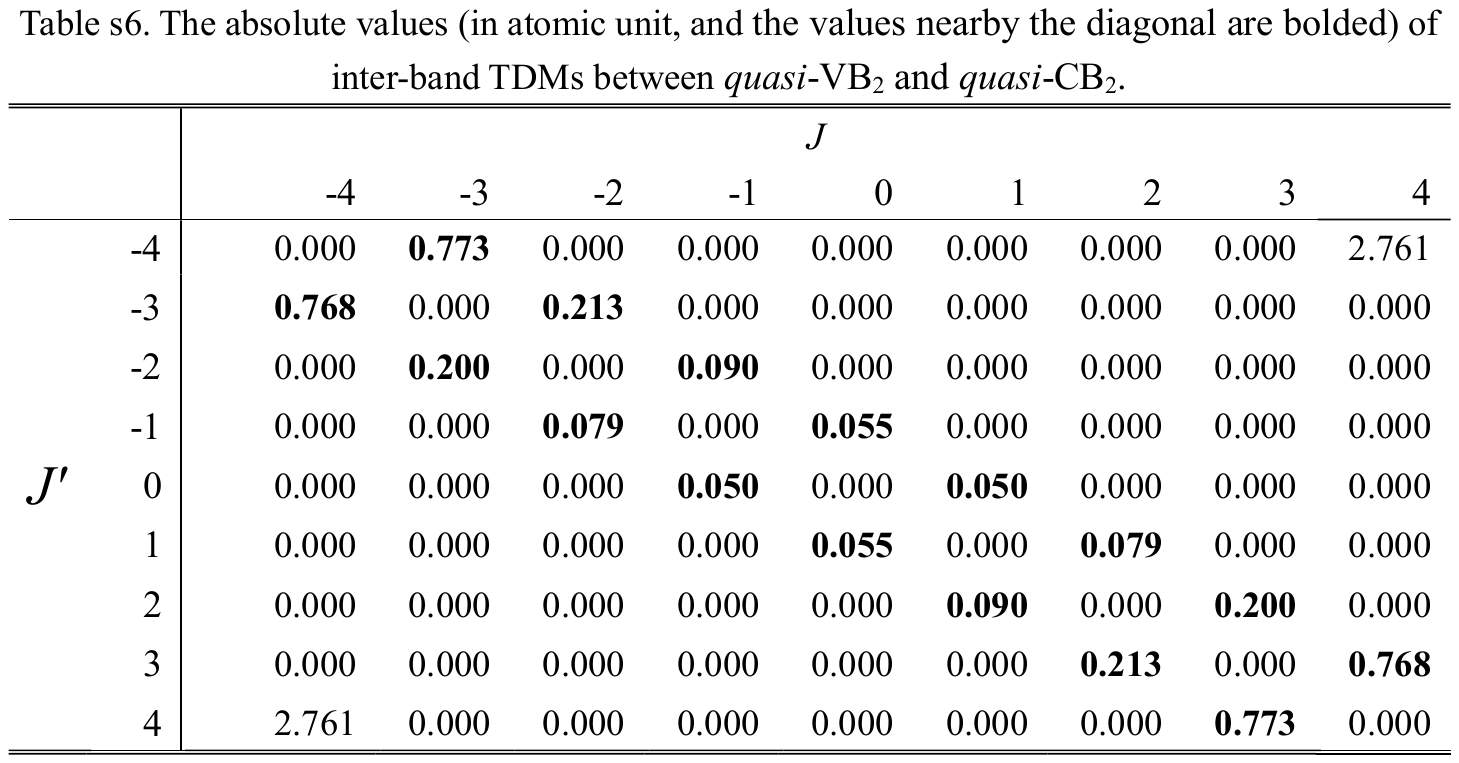}
\includegraphics[width=1.1\textwidth]{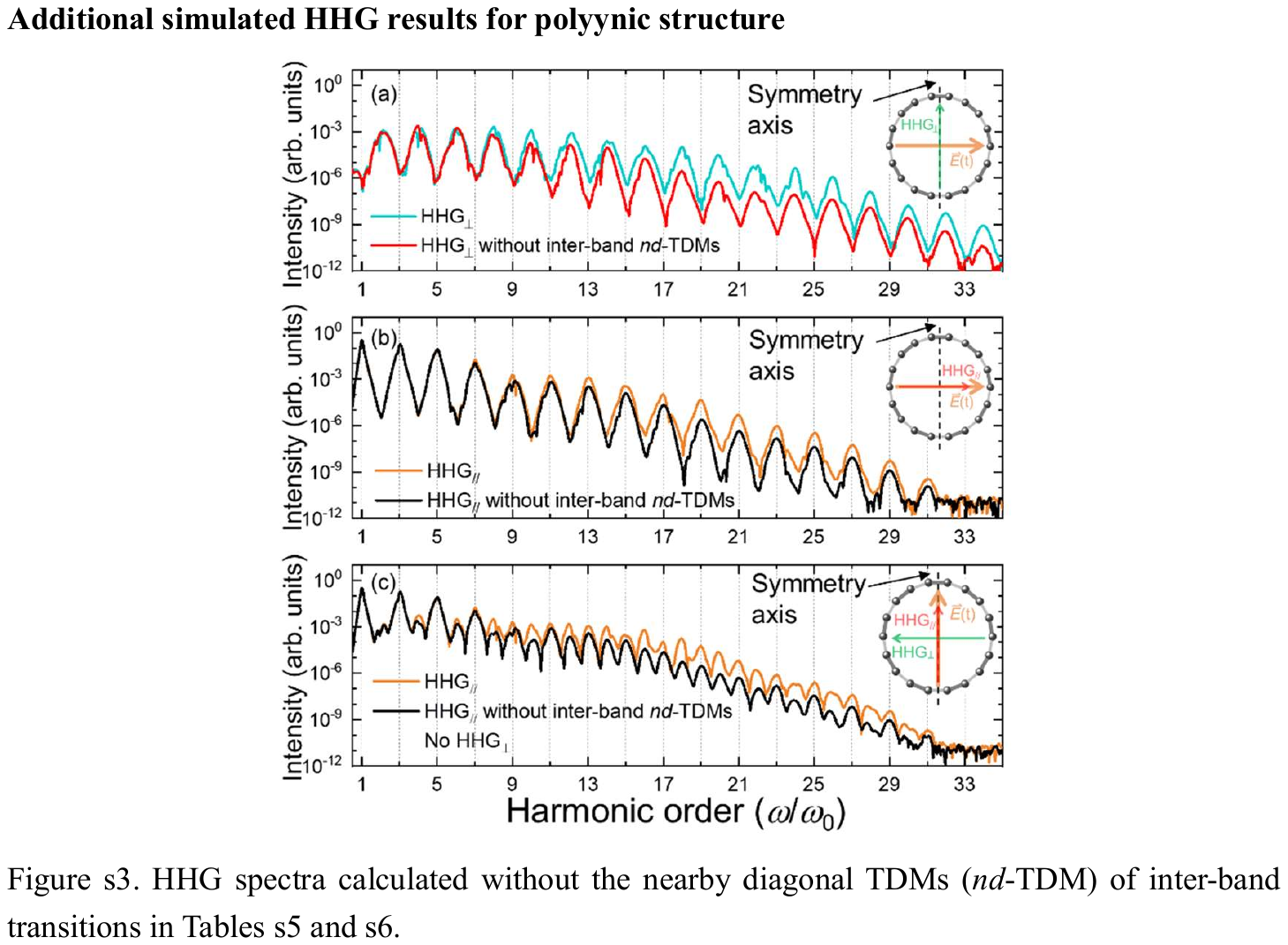}
\includegraphics[width=1.1\textwidth]{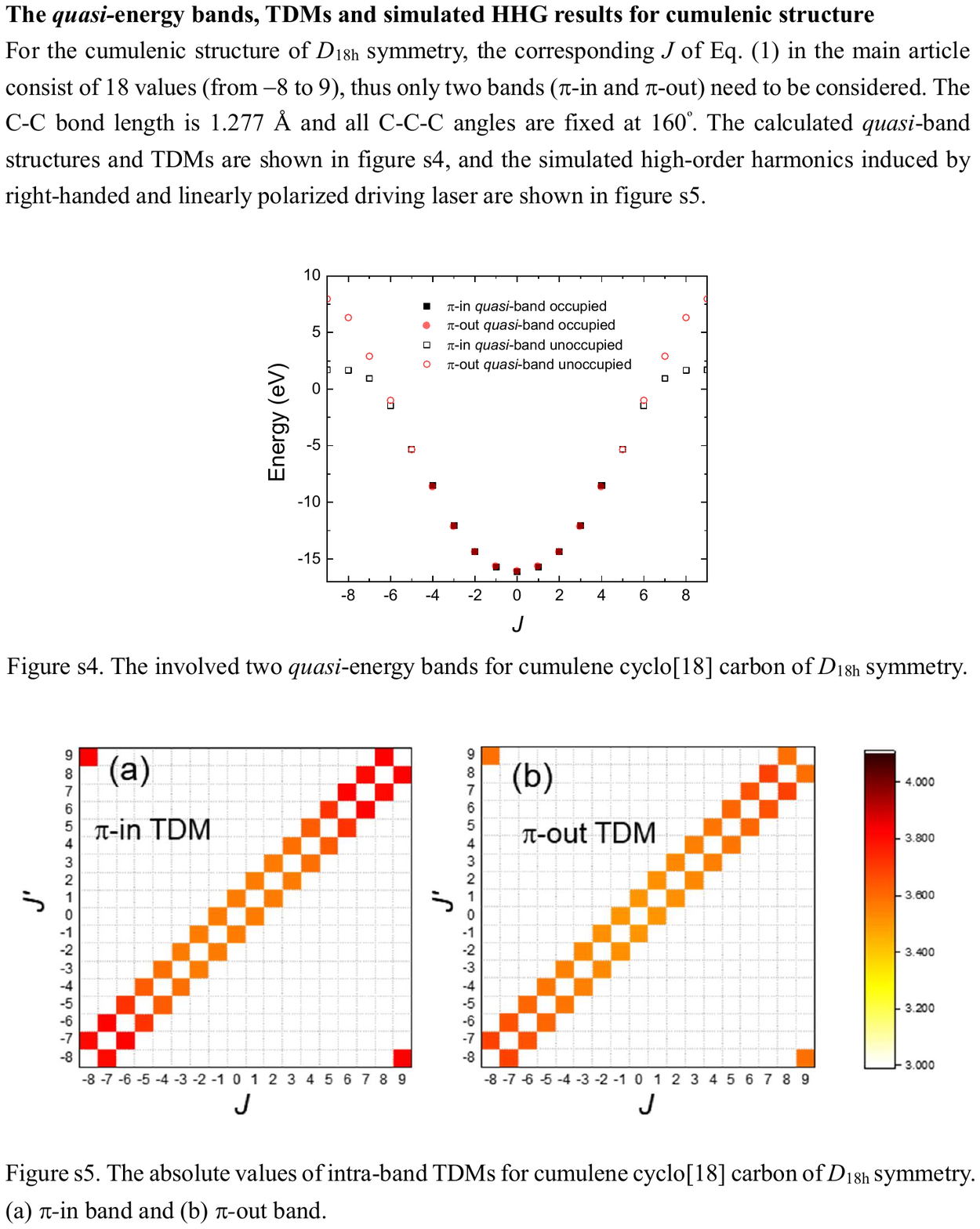}
\includegraphics[width=1.1\textwidth]{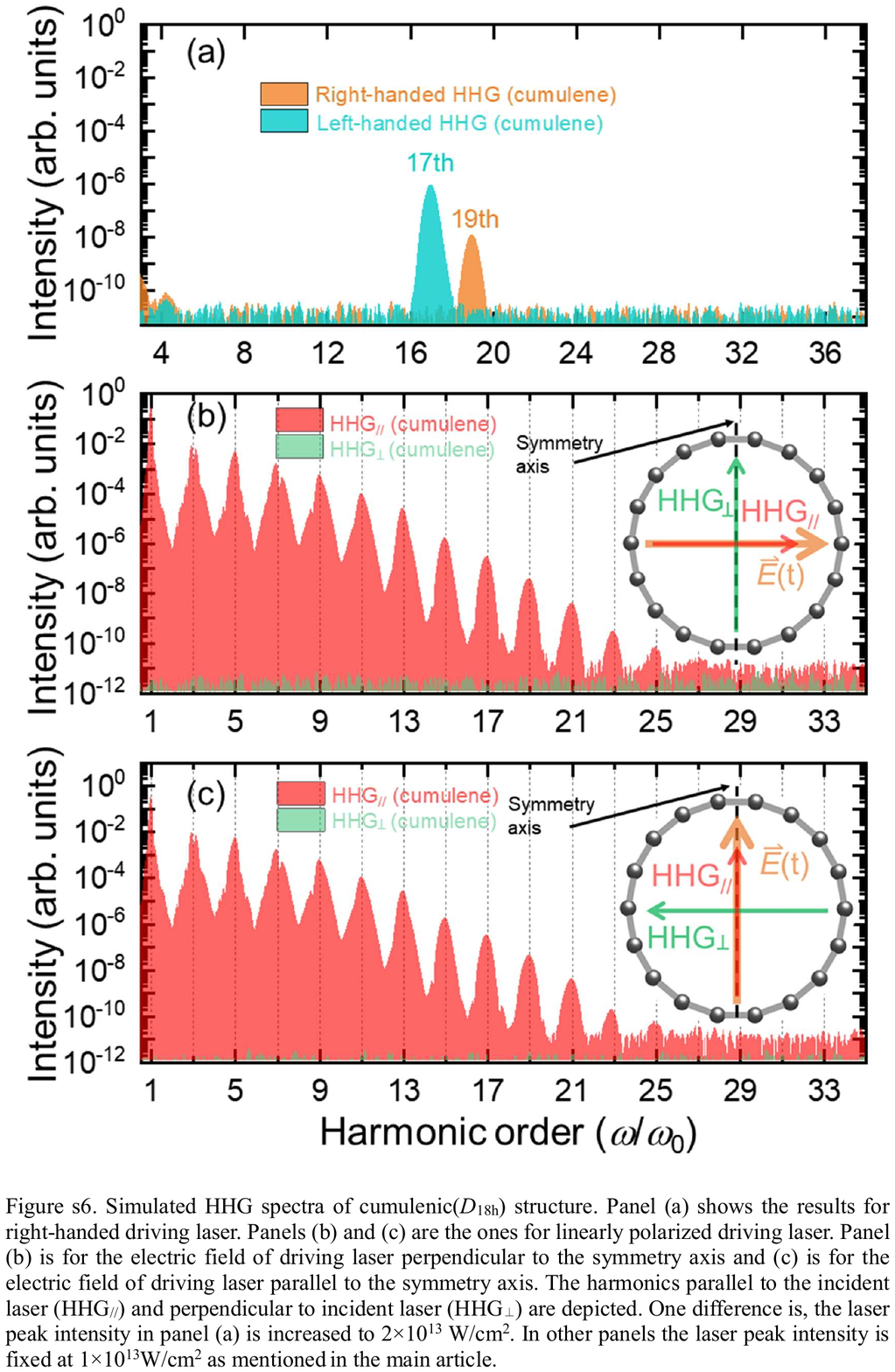}
\end{document}